\documentclass[a4paper,11pt]{article}
\usepackage{pos}
\usepackage{natbib}
\title{From Ground to Space: An Overview of the JEM-EUSO Program for the Study of UHECRs and Astrophysical Neutrinos}
\ShortTitle{An Overview of the JEM-EUSO Program}

\author*[a]{Zbigniew Plebaniak}
\onbehalf{for the JEM-EUSO Collaboration\\[-1mm]{\normalsize \normalfont (a complete list of authors can be found at the end of the proceedings)}
}

\affiliation[a]{Istituto Nazionale di Fisica Nucleare - Sezione di Roma Tor Vergata,\\
 Via della Ricerca Scientifica 1 – 00133, Roma, Italy}

\emailAdd{zbigniew.plebaniak@roma2.infn.it}
\abstract{The JEM-EUSO (Joint Exploratory Missions for Extreme Universe Space Observatory) collaboration is an international initiative studying ultra-high-energy cosmic rays (UHECRs) and related phenomena. These particles, with energies exceeding 10$^{20}$~eV, provide insights into extreme astrophysical processes but remain challenging to detect due to their low flux.

At the heart of JEM-EUSO's technology is an ultra-fast, highly sensitive UV camera capable of detecting extensive air showers in the atmosphere with exceptional spatial and temporal resolution.
In addition, a dedicated Cherenkov camera has been developed to evaluate the viability of the Earth-skimming technique from high altitudes. Fluorescence and Cherenkov detectors can be used together to create a hybrid detection surface, enhancing observational capabilities. This innovative approach enables detailed studies of fluorescence and Cherenkov light from cosmic ray and neutrino interactions. The JEM-EUSO technology will allow for observations from space to significantly increase the exposure to these rare phenomena.

The collaboration employs a multi-platform strategy with ground-based experiments like EUSO-TA calibrating detection systems and validating models, and balloon-borne missions such as EUSO-Balloon and EUSO-SPB1/SPB2 demonstrating observations from the stratosphere and testing advanced technologies. Space-based missions, particularly Mini-EUSO on the ISS, have provided valuable data on UV backgrounds, transient luminous events, and meteoroids, as well as demonstrating the potential for future space-based detection. While we are developing a cross-platform methodology, we are ultimately moving towards space-based measurements. Future efforts include the POEMMA space mission, designed for stereoscopic observations of UHECRs and multi-messenger phenomena, the PBR (POEMMA Balloon with Radio) experiment, which integrates radio detection and is scheduled to fly in 2027, and the M-EUSO satellite mission, proposed to ESA and planned for launch in 2041, if accepted.
Associated experiments also search for meteoroids and strange quark matter (for example, in the form of nuclearites), broadening the scientific scope.
This presentation will summarize the progress of the JEM-EUSO collaboration, highlighting achievements across all platforms and outlining future plans.}

\ConferenceLogo{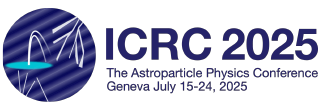}

\FullConference{39th International Cosmic Ray Conference (ICRC2025)\\
 15–24 July 2025\\
Geneva, Switzerland\\}
\begin{document}
\maketitle

\section{Introduction}
\vspace{-0.3em}

Ultra-high-energy cosmic rays (UHECRs) are charged  nuclei from space with energies exceeding 1 EeV. 
Their precise origin and acceleration mechanisms remain unresolved despite decades of study. 
Ground-based observatories have significantly advanced our understanding of their energy spectrum and have provided some insight into their composition, but the extremely low flux of these particles--typically less than one per square kilometer per century--makes statistical analysis difficult, especially at the highest energies.

To address these limitations, the JEM-EUSO program was developed with the goal of observing UHECRs from space, where the available observational area is dramatically larger and the entire celestial sphere can be monitored uniformly. The detection method relies on measuring ultraviolet light produced by extensive air showers (EAS) in the Earth's atmosphere--specifically, the isotropic fluorescence emission from excited nitrogen molecules and the Cherenkov light reflected from the ground or cloud tops. These optical signals allow for the reconstruction of key properties of the primary particle, including its energy, arrival direction, and possibly its identity.
Developing a space-based telescope capable of detecting these signals from low Earth orbit has required the long-term advancement of dedicated optical and electronic technologies, which have been gradually refined over the years. The JEM-EUSO collaboration consists of physicists and engineers from 8 member countries (Czech Republic, France, Germany, Italy, Japan, Poland, Russia, USA) and 4 associated countries (Republic of Korea, Slovakia, Sweden, Switzerland), supported by the world's major space agencies.

JEM-EUSO is a coordinated international program that since 2010 has implemented a series of  experiments to validate and improve the space-based observation technique. These include a ground-based detector EUSO-TA~\cite{Abdellaoui:2018rkw}, stratospheric balloon missions EUSO-Balloon~\cite{Adams:2022oko}, EUSO-SPB1~\cite{JEM-EUSO:2023ypf}, EUSO-SPB2~\cite{Filippatos_SPB2_ICRC2025}, and PBR~\cite{Eser_PBR_ICRC2025} as well as orbital experiments such as TUS onboard the Lomonosov satellite~\cite{Klimov:2017lwx} and Mini-EUSO~\cite{minieuso}, which has been operating on the International Space Station since 2019. 
Recent developments in Cherenkov detection have expanded the scientific reach of EUSO instruments, allowing them to search for high-energy neutrinos by detecting so-called "Earth-skimming" air showers. In addition, the JEM-EUSO instruments are sensitive enough to detect a variety of weak UV phenomena with high time resolution, such as meteors, transient luminous events (e.g., sprites and ELVES), bioluminescence, or even rare exotic signatures like strange quark matter. These secondary science goals demonstrate the broad utility of the JEM-EUSO detection concept.

Ultimately, the knowledge and experience gained through the JEM-EUSO program are laying the groundwork for future large-scale missions such as K-EUSO~\cite{Klimov:2022jzk}, POEMMA~\cite{POEMMA:2020ykm} and M-EUSO. These next-generation satellites are being designed to achieve the exposure and sensitivity necessary to finally resolve the fundamental questions surrounding the origin and nature of the most energetic particles in the Universe.
In this work, we review the JEM-EUSO
%related experiments carried out so far and 
program and present the current perspectives for future space-based UHECR missions.

\begin{figure}[t]
\centering
\hfill
\includegraphics[width=1.0\textwidth]{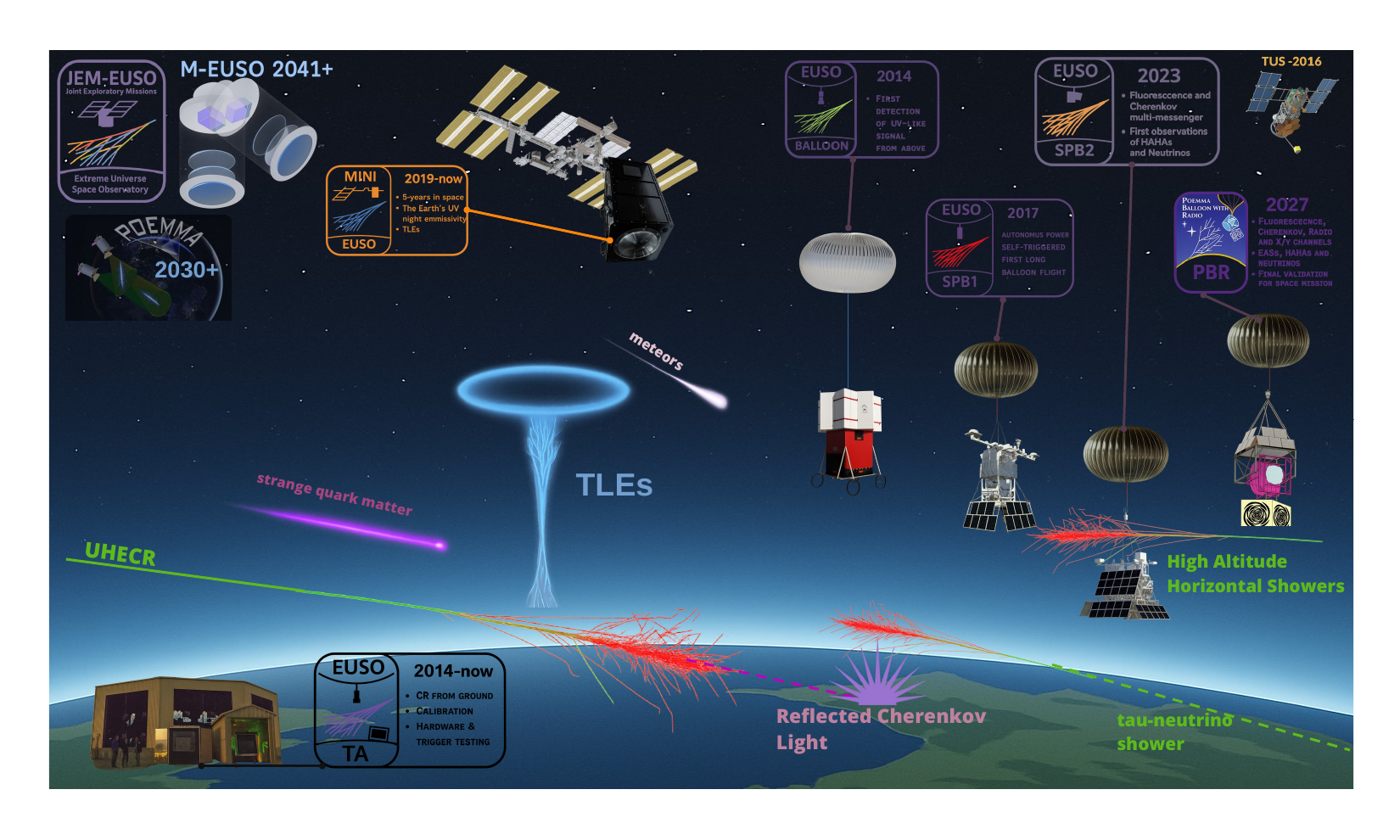}
%}
%\hfill
\vspace{-1.1em}
\caption{Scientific objectives of the various JEM-EUSO missions. For detailed descriptions, see the main text.}

\label{fig:MiniEUSOData}
\end{figure}

\section{EUSO-TA: A Ground-Based EUSO Laboratory}

EUSO-TA is a ground-based fluorescence telescope operating since 2014 at the Black Rock Mesa site of the Telescope Array (TA) experiment in Utah, USA~\cite{TAMEDA200974}. Designed as a pathfinder for space-based UHECR observatories, it detects ultraviolet fluorescence light emitted by extensive air showers induced by high-energy cosmic rays. The EUSO-TA detector features a dual Fresnel lens system and a focal surface comprising a single Photo-Detector Module (PDM) with 36 Multi-Anode Photomultiplier Tubes (MAPMTs), totaling 2304 pixels. The telescope provides a field of view of $10.5^\circ \times 10.5^\circ$, a spatial resolution of about $0.2^\circ$ per pixel, and a temporal resolution of $2.5\ \mu\mathrm{s}$.
EUSO-TA operates in parallel with the TA experiment and can be externally triggered by TA events, enabling detailed cross-checks of event reconstruction and timing. Across several observation campaigns, it has recorded nine TA-triggered UHECR events. Calibration procedures include laser shots from the Central Laser Facility, mobile GLS systems, and stellar tracking~\cite{Plebaniak:2023mhu}.
The EUSO-TA site serves as a long-term validation platform for instruments developed within the JEM-EUSO collaboration, including those for future balloon and space missions. Since 2022, the detector has been undergoing an upgrade, adopting electronics and acquisition systems similar to those of Mini-EUSO. Further developments focus on automating its operation.

\section{Stratospheric Balloon Program}
%\section{Balloon - borne experiments}

Stratospheric balloon experiments offer a unique opportunity to develop and test EUSO technology in near-space conditions. Operating above 30~km altitude exposes the instruments to a harsh environment, particularly challenging for the high-voltage power systems, sometimes even more so than in space. So far, three balloon missions have been carried out within the JEM-EUSO program, with the fourth one currently in preparation.

\begin{figure}[t]
\centering
\hfill
\includegraphics[width=0.42\textwidth]{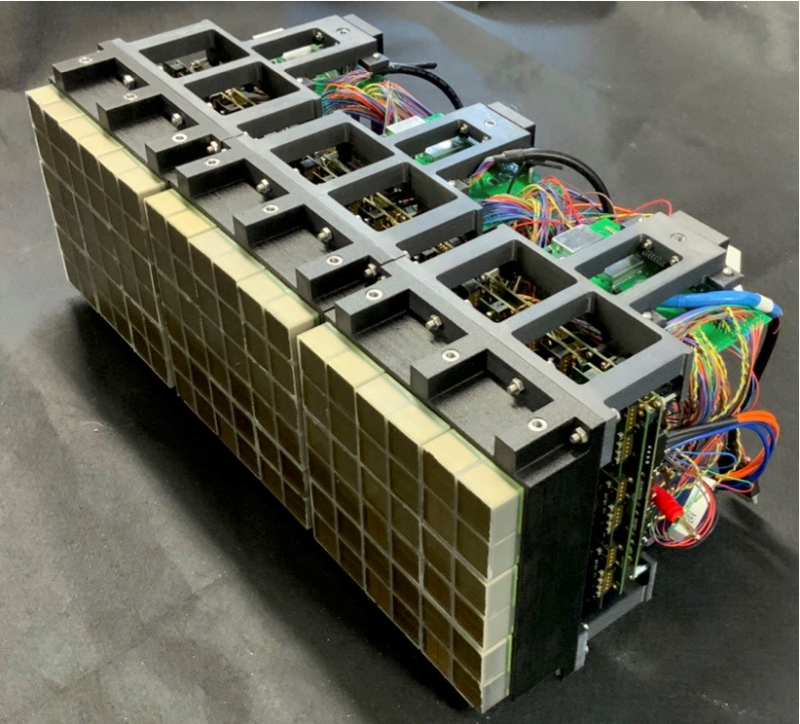}
%\hfill
%\raisebox{1.3em}{
\includegraphics[width=0.57\textwidth]{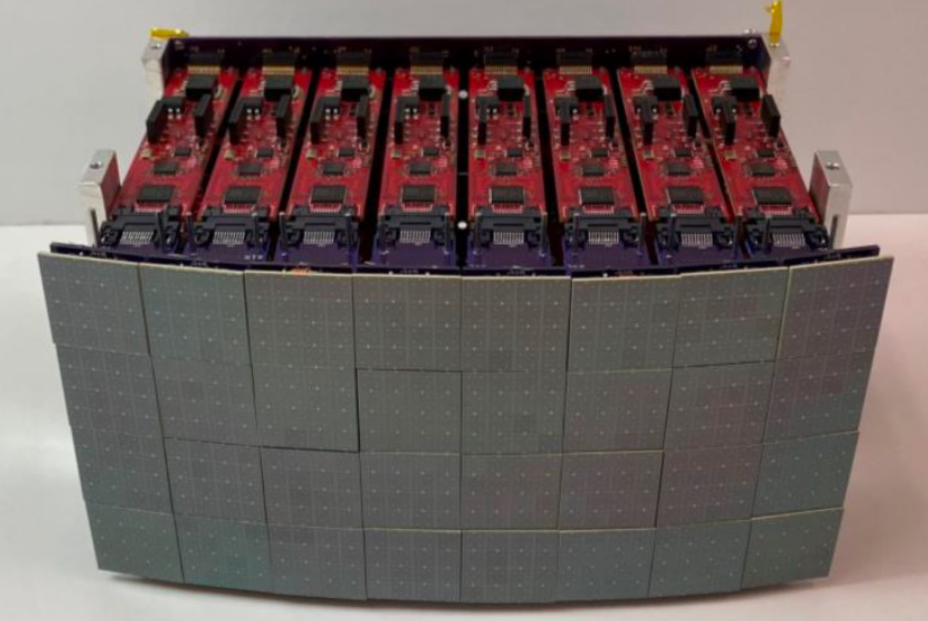}
%}
%\hfill
\vspace{-1.1em}
\caption{Focal surfaces of the latest EUSO detectors from EUSO-SPB2 mission. \textbf{Left:} Fluorescence camera composed of three PDMs, including together 108 multi-anode photomultiplier tubes, providing a total of 6912 pixels. The data are read out with a time resolution of 1~$\mu$s. \textbf{Right:} Cherenkov camera consisting of 512 pixels based on silicon photomultipliers (SiPMs), operating with a time resolution of 10~ns.}
\label{fig:MiniEUSOData}
\end{figure}

%\begin{itemize} 
    %\item 
    {\bf EUSO-Balloon}~\cite{Adams:2022oko} was launched by the CNES from the Timmins base in Ontario, Canada, on August 24th, 2014. 
    Observing the ground for approximately 5 hours in nadir orientation, the detector equipped with one PDM and dedicated ASIC front-end electronics measured the UV emissivity of the Earth's surface. The optical system consisted of two Fresnel lenses with a total area of about 1\,m$^2$ and a field of view of $6^\circ \times 6^\circ$. 
    Operating in continuous data-taking mode, EUSO-Balloon successfully detected laser light tracks and xenon flashers emitted from a helicopter flying below the balloon, demonstrating the feasibility of observing EAS-like events with an EUSO-type detector.

    %\item 
    {\bf EUSO-SPB1} was a mission of opportunity flown aboard a NASA super-pressure balloon, launched from the NASA Balloon Base in Wanaka, New Zealand, in April 2017, and originally intended for a multi-week flight around the Southern Hemisphere. Its primary scientific objective was to perform the first observation of EAS induced by UHECRs from 33\,km altitude, using a nadir-pointing UV camera. The detector was equipped with an optical system similar to that of EUSO-Balloon and a second-generation front-end electronics chain (an ASIC based on SPACIROC3) to read out signals from a single PDM. The flight was terminated after 12 days due to a leak in the balloon. Although EUSO-SPB1 did not record any UHECR-like events, the detector operated nominally throughout the flight, collecting 25.1 hours of data and contributing to the increase of the Technology Readiness Level (TRL) of the EUSO fluorescence camera.

    {\bf EUSO-SPB2} was launched from Wanaka in May 2023 for a flight originally expected to last several weeks, similar to the SPB1 mission. For the first time, the payload included two independent optical telescopes. The fluorescence telescope, equipped with 3 PDMs (6912 pixels), used a 1-meter Schmidt optical system with a field of view of $36^\circ \times 12^\circ$, designed to detect EAS tracks with 1\,$\mu$s time resolution. The Cherenkov telescope, with 512 SiPM pixels, was pointed $10^\circ$ below the Earth’s limb to search for direct Cherenkov flashes from high-energy neutrinos, achieving 10\,ns resolution.
    Due to a leak in the balloon envelope, the flight was terminated after about 37 hours. No UHECR events were found among the 120,000 recorded fluorescence triggers, but several neutrino candidates were identified in the 32,000 downloaded bifocal Cherenkov triggers. The mission successfully validated the new trigger scheme and demonstrated the feasibility of the Cherenkov-from-space detection technique.

{\bf The POEMMA Balloon with Radio (PBR)}, currently under development, is a pathfinder mission based on the POEMMA design study and previous EUSO experiments. It is planned to follow a similar super-pressure balloon flight profile as EUSO-SPB1 and EUSO-SPB2, with launch scheduled for spring 2027.

The instrument will feature an upgraded Schmidt optical system with a 1.1\,m entrance pupil and two improved detectors: a Fluorescence Camera (FC) and a Cherenkov Camera (CC). The FC, with a $24^\circ \times 24^\circ$ field of view, will use 4 PDMs (9216 pixels), providing $\sim$$0.2^\circ$ angular resolution and 1\,$\mu$s time resolution. The CC, with a $12^\circ \times 6^\circ$ field of view and 2048 SiPM pixels, will also reach $0.2^\circ$ angular resolution and operate at 200\,MHz sampling. A bifocal optical system will reduce noise by requiring coincident spots to trigger events.
Additionally, the Radio Instrument (RI) will use two dual-polarized sinuous antennas to detect EAS radio signals in the 50–500\,MHz range, complementing optical Cherenkov detection and improving event reconstruction, especially for high-altitude horizontal air showers (HAHAs), while enhancing sensitivity to tau neutrinos.

PBR has three main scientific goals: (1) detect UHECR fluorescence light from suborbital altitudes; (2) study HAHA events using Cherenkov and radio data for better spectrum and composition analysis near the PeV scale; and (3) search for astrophysical neutrinos, especially Earth-skimming tau neutrinos, in response to transient alerts. The mission will raise the TRL of the key systems for the future POEMMA and M-EUSO missions.

\vspace{-0.1cm}
\section{Mini-EUSO: A Space-borne EUSO Mission}

{\bf Mini-EUSO} (Multiwavelength Imaging New Instrument for the Extreme Universe Space Observatory) is a compact UV telescope operating aboard the International Space Station since 2019. It was designed to measure photon rates per pixel from diffuse UV sources under conditions analogous to those expected for future large-scale space missions, using a larger pixel FoV to compensate for its smaller optics limited by the ISS window size~\cite{Bertaina:icrc2025}.
The instrument uses a Fresnel-lens optical system and a focal surface with 2304 channels (1~PDM) to observe the night-time Earth in the 290--430~nm range, with 6.3~km spatial and 2.5~$\mu$s temporal resolution, through a nadir-facing UV-transparent window in the Zvezda module. It operates at three integration scales: 2.5~$\mu$s, 320~$\mu$s, and 40.96~ms which allow for observations of various atmospheric phenomena, such as transient luminous events (TLEs), and contributed to the search for strange quark matter as a dark matter candidate. The intensity as a function of signal duration for different classes of events is shown in Fig.~\ref{fig:MiniEUSOData} (left).
One of the main scientific goals of the Mini-EUSO mission was to produce a night-time UV map of the Earth, see Fig.~\ref{fig:MiniEUSOData} (right). Over 5 years of observations, Mini-EUSO built the largest space-based meteor database to date.  Additionally, it validated trigger schemes for future space-based UHECR detectors.

\begin{figure}[t]
\centering
\hfill
\includegraphics[width=0.55\textwidth]{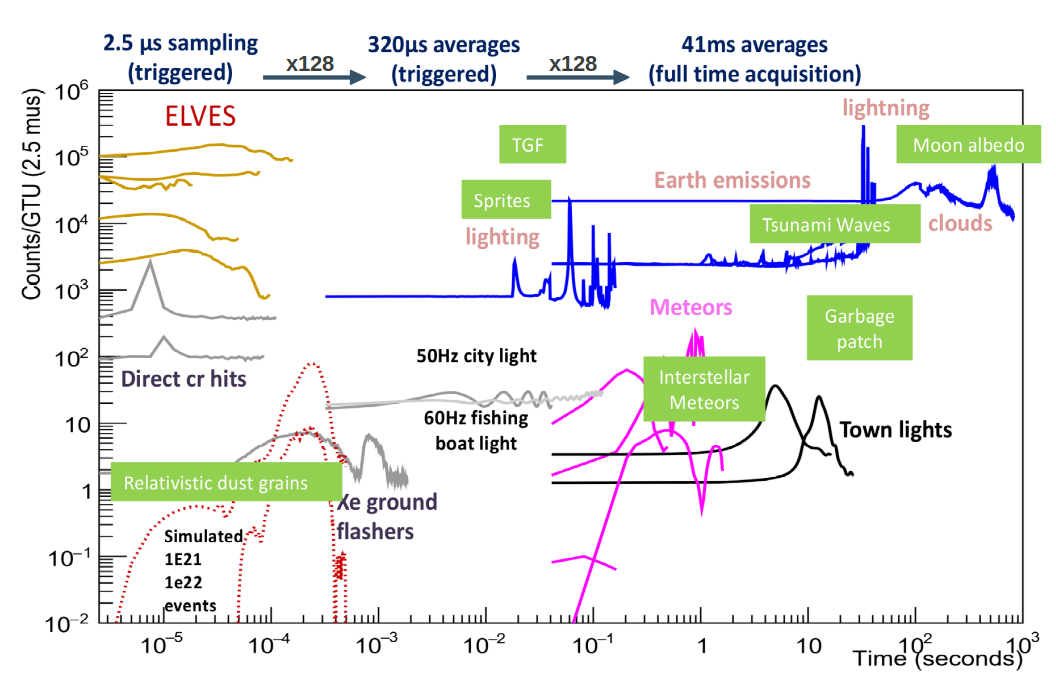}
%\hfill
\raisebox{1.3em}{\includegraphics[width=0.42\textwidth]{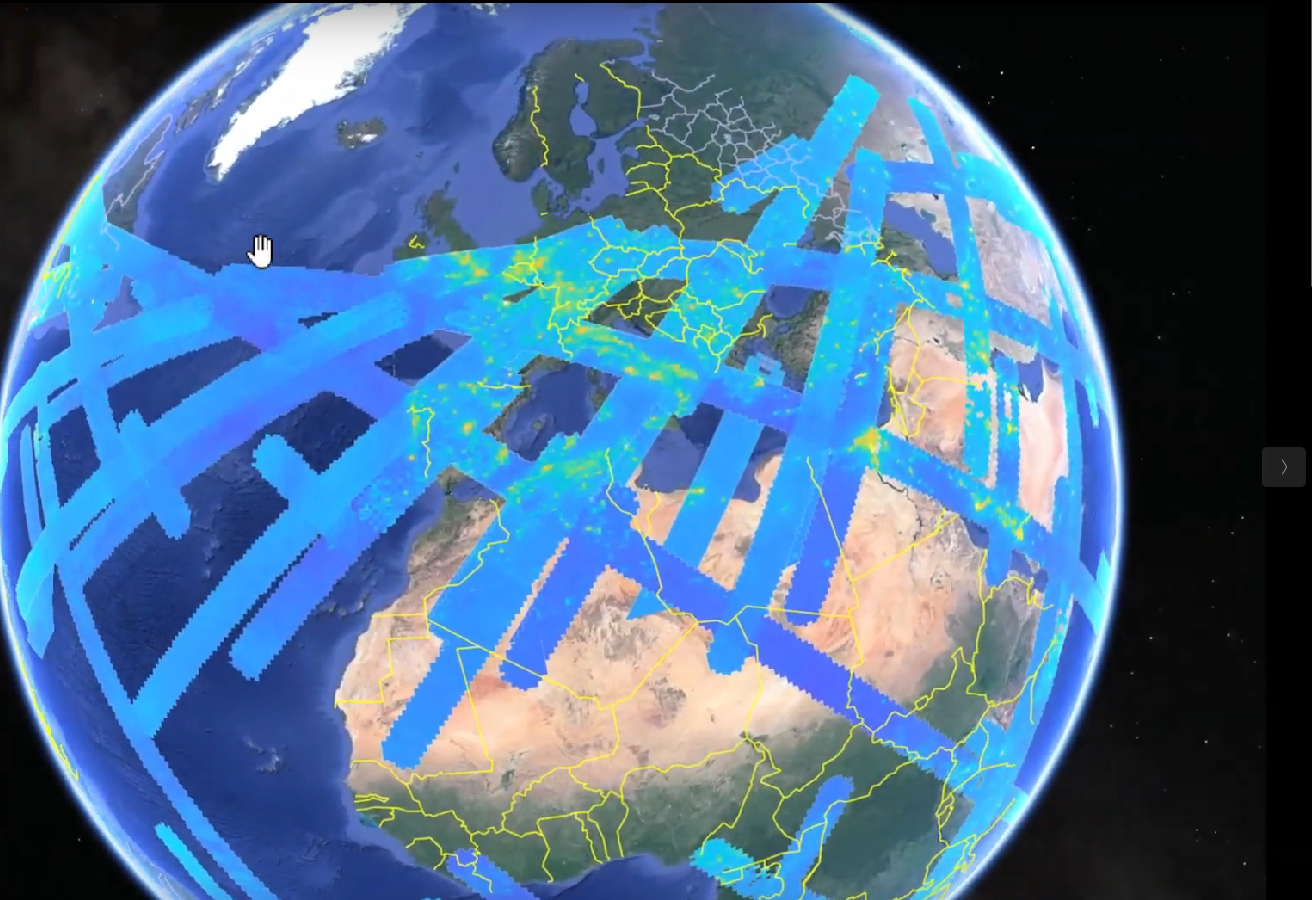}}
\hfill
%\vspace{-0.1em}
\caption{%Variety of signals measured by Mini-EUSO. 
\textbf{Left:} Temporal profiles of different events. Real signals measured at three time scales are shown alongside a simulated UHECR profile. \textbf{Right:} Partial map of the Earth's UV night sky reconstructed from Mini-EUSO data.}
\label{fig:MiniEUSOData}
\end{figure}

\section{Future Missions and the M-EUSO Project}

Building on experience from the JEM-EUSO collaboration, we conducted design studies for several satellite-based missions, including K-EUSO, previously considered as an ISS-attached instrument, and POEMMA, a NASA concept with two satellites flying in stereo configuration, planned for the next decade. These efforts led to the proposal of the new M-EUSO satellite mission in response to the ESA M8 call, with a planned flight date in 2041~\cite{ESA_M8_Call_2025}.

\textbf{M-EUSO} (Multi-messenger Extreme Universe Space Observatory) will represent the next major step in the JEM-EUSO program, designed to probe the highest-energy phenomena in the Universe. It will be a dedicated ESA-led space-based observatory aiming to detect UHECRs and astrophysical neutrinos from low Earth orbit using ultraviolet (UV)  air-shower imaging. Operating as a free-flyer with near-uniform sky coverage, M-EUSO will enable the first statistically robust full-sky map of UHECRs and support multi-messenger astronomy at extreme energies.

\subsection{Instrument and Mission Design}

M-EUSO will consist of a single large-aperture UV telescope with a dual-detector focal surface, optimized for detecting both fluorescence and Cherenkov signals from extensive air showers.

The telescope will use a 2.3-meter diameter Fresnel lenses providing a square-equivalent field of view of $\sim$40$^\circ\times$40$^\circ$ to focus UV light. The focal surface will include two detectors:
\begin{itemize}
\item \textbf{Fluorescence Channel:} Light will be focused onto 49 multi-anode PMT modules ($\sim$113k pixels), offering single-photon sensitivity with 2.5~$\mu$s sampling for UHECR reconstruction and longer integration times for complementary science.

\item \textbf{Cherenkov Channel:} Located at the FoV edge and containing $\sim$6,000 SiPM pixels with time sampling at 10~ns, it will detect fast beamed Cherenkov light from near-horizontal showers.
\end{itemize}

A separate infrared (IR) camera will monitor cloud cover to correct atmospheric effects and improve data quality. The telescope will operate from a $\sim$450 km orbit with 30$^\circ$ inclination. Two observational modes are foreseen:
\begin{itemize}
\item \textbf{Nadir mode} (initial phase): Optimized for UHECR detection with high resolution in energy and direction. The annual exposure will be $\geq$30,000 km$^2\cdot$sr$\cdot$yr, exceeding in one year the lifetime exposure of the Telescope Array and reaching about 25\% of PAO's 20-year exposure.
\item \textbf{Tilted mode} (second phase): Will expand geometrical aperture, enabling detection of neutrinos and transients. This mode will reach $\sim$100,000 km$^2\cdot$sr$\cdot$yr per year above 3$\cdot$10$^{20}$ eV and enable observation of upward-going air showers from Earth-skimming $\nu_\tau$ interactions.
\end{itemize}

\textbf{Expected Performance}

M-EUSO will detect hundreds of UHECRs per year above 30 EeV, achieving full efficiency beyond 90 EeV with angular resolution better than 3$^\circ$. For neutrinos, it will complement ground detectors at higher energies and broader sky coverage. Simulations show sensitivity to $\nu_\tau$ fluxes above 40 PeV.

\subsection{Scientific Objectives of M-EUSO}

M-EUSO will open a new window onto the extreme Universe by detecting ultra-high-energy cosmic rays and very high-energy neutrinos from space. Its main objectives are:

\begin{itemize}

\item \textbf{UHECR Source Identification:} Reconstruct the spectrum, anisotropy, and arrival directions of cosmic rays above 30 EeV. M-EUSO will (1) create the first full-sky map of UHECRs with uniform exposure to detect anisotropies and active regions, (2) measure the energy spectrum at the highest energies to constrain acceleration and propagation models, and (3) resolve differences between Northern and Southern hemisphere observations through consistent, global data.
\item \textbf{Neutrino Astronomy:} Detect Earth-skimming $\nu_\tau$ ($>$40 PeV) using fast Cherenkov imaging to study extreme transients.
\item \textbf{Fundamental Physics:} Probe hadronic interactions above LHC energies and search for new physics such as Lorentz invariance violation or super-heavy dark matter decay.
\item \textbf{Multi-Messenger Follow-Up:} Respond to alerts from gravitational wave, gamma-ray, and neutrino observatories.
\end{itemize}

M-EUSO will also support interdisciplinary science, including studies of transient luminous events, meteors, auroras, atmospheric disturbances, and rare phenomena like nuclearites (strange quark matter) and exotic UV flashes.

\subsection{Technological Readiness and Role in ESA's Multi-Messenger Strategy}

M-EUSO will be built on over 20 years of development in the JEM-EUSO collaboration. Its technologies have been validated in several pathfinder missions (Mini-EUSO, EUSO-SPB1/SPB2, PBR), ensuring high technological readiness (TRL $\geq$ 6).

The mission will align with ESA's Voyage 2050 strategy in multi-messenger astronomy, complementing missions like Athena (X-rays) and LISA (gravitational waves) by targeting UHECRs and neutrinos. M-EUSO will employ proven, flight-ready technologies and its flexible design will support a broad range of science goals, including planetary, atmospheric, and exotic physics studies.

\section*{Acknowledgements}
\small
The authors would like to acknowledge the support by
NASA award 11-APRA-0058, 
16-APROBES16-0023, 17-APRA17-0066, 
NNX17AJ82G, NNX13AH54G, 
80NSSC18K0246, 80NSSC18K0473,\\ 80NSSC19K0626,
80NSSC18K0464, 
80NSSC22K1488, 80NSSC19K0627 and 
80NSSC22K0426, by the French space agency CNES, This research used resources of the National Energy Research Scientific Computing Center (NERSC), a U.S. Department of Energy Office of Science User Facility operated under Contract No. DE-AC02-05CH11231. We acknowledge the ASI-INFN agreement n. 2021-8-HH.0, its amendments and agreement n. 2020-26-Hh.0. We acknowledge the NASA Balloon Program Office and the Columbia Scientific Balloon Facility and staff for extensive support. The Mini-EUSO, K-EUSO and TUS activities have also been supported by the state assignment of Lomonosov Moscow State University and by the Russian State Space Corporation Roscosmos.

\bibliography{my-bib-database}

@article{Adams:2022oko,
    author = "Adams, J. H. and others",
    title = "{A Review of the EUSO-Balloon Pathfinder for the JEM-EUSO Program}",
    doi = "10.1007/s11214-022-00870-x",
    journal = "Space Sci. Rev.",
    volume = "218",
    number = "1",
    pages = "3",
    year = "2022"
}

@article{Abdellaoui:2018rkw,
    author = "Abdellaoui, G. and others",
    title = "{EUSO-TA \textendash{} First results from a ground-based EUSO telescope}",
    doi = "10.1016/j.astropartphys.2018.05.007",
    journal = "Astropart. Phys.",
    volume = "102",
    pages = "98--111",
    year = "2018"
}

@article{JEM-EUSO:2023ypf,
    author = "Abdellaoui, G. and others",
    collaboration = "JEM-EUSO",
    title = "{EUSO-SPB1 mission and science}",
    eprint = "2401.06525",
    archivePrefix = "arXiv",
    primaryClass = "astro-ph.IM",
    doi = "10.1016/j.astropartphys.2023.102891",
    journal = "Astropart. Phys.",
    volume = "154",
    pages = "102891",
    year = "2024"
}

@article{Klimov:2017lwx,
    author = "Klimov, P. A. and others",
    title = "{The TUS detector of extreme energy cosmic rays on board the Lomonosov satellite}",
    eprint = "1706.04976",
    archivePrefix = "arXiv",
    primaryClass = "astro-ph.IM",
    doi = "10.1007/s11214-017-0403-3",
    journal = "Space Sci. Rev.",
    volume = "212",
    number = "3-4",
    pages = "1687--1703",
    year = "2017"
}

@misc{ESA_M8_Call_2025,
  author       = "{European Space Agency (ESA)}",
  title        = "{{CALL FOR A MEDIUM-SIZE AND A FAST MISSION OPPORTUNITY (M8 \& F3)}}",
  howpublished = "ESA Call M8 documentation",
  month        = mar,
  year         = 2025,
  note         = "Solicits proposals for M‑class missions; includes JEM‑EUSO M‑EUSO proposal",
  url          = {https://www.cosmos.esa.int/web/call-for-missions-2025},
}

@article{minieuso,
    author = "Bacholle, S. and others",
    title = "{Mini-EUSO Mission to Study Earth UV Emissions on board the ISS}",
    eprint = "2010.01937",
    archivePrefix = "arXiv",
    primaryClass = "astro-ph.IM",
    doi = "10.3847/1538-4365/abd93d",
    journal = "Astrophys. J. Suppl.",
    volume = "253",
    number = "2",
    pages = "36",
    year = "2021"
}

@article{Klimov:2022jzk,
    author = "Klimov, Pavel and others",
    title = "{Status of the K-EUSO Orbital Detector of Ultra-High Energy Cosmic Rays}",
    eprint = "2201.12766",
    archivePrefix = "arXiv",
    primaryClass = "astro-ph.IM",
    doi = "10.3390/universe8020088",
    journal = "Universe",
    volume = "8",
    number = "2",
    pages = "88",
    year = "2022"
}

@article{POEMMA:2020ykm,
    author = "Olinto, A. V. and others",
    collaboration = "POEMMA",
    title = "{The POEMMA (Probe of Extreme Multi-Messenger Astrophysics) observatory}",
    eprint = "2012.07945",
    archivePrefix = "arXiv",
    primaryClass = "astro-ph.IM",
    doi = "10.1088/1475-7516/2021/06/007",
    journal = "JCAP",
    volume = "06",
    pages = "007",
    year = "2021"
}

@article{TAMEDA200974,
title = {{Telescope Array} Experiment},
journal = {Nuclear Physics B - Proceedings Supplements},
volume = {196},
pages = {74-79},
year = {2009},
note = {Proceedings of the XV International Symposium on Very High Energy Cosmic Ray Interactions (ISVHECRI 2008)},
issn = {0920-5632},
doi = {https://doi.org/10.1016/j.nuclphysbps.2009.09.011},
url = {https://www.sciencedirect.com/science/article/pii/S0920563209006471},
author = {Y. Tameda}
}

@article{Plebaniak:2023mhu,
    author = "Plebaniak, Zbigniew Dariusz and Przybylak, Marika and others", 
    collaboration = "JEM-EUSO",
    title = "{Calibration and testing of the JEM-EUSO detectors using stars observed in the UV band}",
    journal = "PoS",
    volume = "ICRC2023",
    pages = "386",
    year = "2023"
}

@article{Filippatos_SPB2_ICRC2025,
    author = "Filippatos, G. and others", 
    collaboration = "JEM-EUSO",
    title = "{EUSO-SPB2 Cosmic Ray Searches and Observations}",
    journal = "PoS",
    volume = "ICRC2025",
    pages = "255",
    year = "2025"
}

@article{Eser_PBR_ICRC2025,
    author = "Eser, J. and others", 
    collaboration = "JEM-EUSO",
    title = "{POEMMA-Balloon with Radio: An Overview}",
    journal = "PoS",
    volume = "ICRC2025",
    pages = "249",
    year = "2025"
}

@article{Bertaina:icrc2025,
    author = "Bertaina, M.E.  and others", 
    collaboration = "JEM-EUSO",
    title = "{Implications of Mini-EUSO measurements for a space-based observation of UHECRs}",
    journal = "PoS",
    volume = "ICRC2025",
    pages = "189",
    year = "2025"
}

    \newpage
{\Large\bf Full Authors list: The JEM-EUSO Collaboration}
%{\scriptsize (author-list as of September 15th, 2025 )} \hspace{0.6cm}
%{\scriptsize (version  \today{} \currenttime{})}
%\vspace*{0.5cm}
%contact: zbigniew.plebaniak@roma2.infn.it, marco.ricci@lnf.infn.it

\begin{sloppypar}
{\small \noindent
M.~Abdullahi$^{ep,er}$              % Italy
M.~Abrate$^{ek,el}$,                % Italy
J.H.~Adams Jr.$^{ld}$,              % USA 
D.~Allard$^{cb}$,                   % France
P.~Alldredge$^{ld}$,                % USA
R.~Aloisio$^{ep,er}$,               % Italy
R.~Ammendola$^{ei}$,                % Italy
A.~Anastasio$^{ef}$,                % Italy %%
L.~Anchordoqui$^{le}$,              % USA
V.~Andreoli$^{ek,el}$
% ...
}
\end{sloppypar}
\vspace*{.3cm}

%%\newpage
{ \footnotesize
\noindent
%
% Czech Republic - 2 intitutions
%Czech Republic\\
% ...
$^{cb}$ Universit\'e de Paris, AstroParticule et Cosmologie (CNRS), Paris, France\\
% ...
$^{ef}$ Istituto Nazionale di Fisica Nucleare (INFN), Sezione di Napoli, Naples, Italy\\
% ...
$^{ei}$ Istituto Nazionale di Fisica Nucleare (INFN), Sezione di Roma Tor Vergata, Rome, Italy\\
% ...
$^{ek}$ Istituto Nazionale di Fisica Nucleare (INFN), Sezione di Torino, Turin, Italy\\
$^{el}$ Universit\`a di Torino, Dipartimento di Fisica, Turin, Italy\\
% ...
$^{ep}$ Gran Sasso Science Institute (GSSI), L’Aquila, Italy\\
$^{er}$ Istituto Nazionale di Fisica Nucleare (INFN), Laboratori Nazionali del Gran Sasso, Assergi, Italy\\
% ...
$^{ld}$ University of Alabama in Huntsville, Huntsville, AL, USA\\
$^{le}$ City University of New York (CUNY), Lehman College, Bronx, NY, USA\\
% ...
}

%\begin{thebibliography}{99}
%\bibitem{...}
%....

%\end{thebibliography}

\end{document}